\pdfoutput=1

\documentclass[british,english,american]{article}
\usepackage{charter}
\usepackage[T1]{fontenc}
\usepackage[latin9]{inputenc}
\usepackage[a4paper]{geometry}
\usepackage{babel}
\usepackage{float}
\usepackage{amsmath}
\usepackage{graphicx}
\usepackage{esint}
\usepackage[unicode=true,pdfusetitle,
 bookmarks=true,bookmarksnumbered=false,bookmarksopen=false,
 breaklinks=false,pdfborder={0 0 0},backref=false,colorlinks=false,
 pdfkeywords={Electromagnetic optics, Polarization, Statistical optics, Coherence, Noise in imaging systems}]
 {hyperref}
\usepackage{breakurl}

\usepackage{multicol}
\usepackage{cite}
\usepackage{amssymb}

\allowdisplaybreaks[1] % 1=allow 2,3,4=increasingly permissive

\begin{document}

\title{Image quality in double- and triple-intensity ghost imaging \\
with classical partially polarized light}

\author{Henri Kellock,$^{1,*}$ Tero Setälä,$^{1}$ Tomohiro Shirai,$^{2}$
and Ari T. Friberg$^{1,3,4}$ }
\date{\vspace{-5ex}}
\maketitle

{\centering\small
  \noindent{$^{1}$\emph{Department of Applied Physics, Aalto University, P. O. Box 13500, FI-00076 Aalto, Finland}}

  \noindent{$^{2}$\emph{Electronics and Photonics Research Institute,  National Institute of Advanced Industrial Science and Technology, AIST Tsukuba Central 2, 1-1-1 Umezono, Tsukuba 305-8568, Japan}}

  \noindent{$^{3}$\emph{Department of Physics and Mathematics, University of Eastern Finland, P. O. Box 111,
  FI-80101 Joensuu, Finland}}

  \noindent{$^{4}$\emph{Department of Microelectronics and Applied Physics, Royal Institute of Technology, Electrum 229,
  SE-164 40 Kista, Sweden}}

  \noindent{$^{*}$\emph{Corresponding author:} henri.kellock@aalto.fi}
  
  \noindent{\today}
  
}

\begin{abstract}
Classical ghost imaging is a correlation-imaging technique in which
the image of the object is found through intensity correlations of
light. We analyze three different quality parameters, namely the visibility,
the signal-to-noise ratio (SNR), and the contrast-to-noise ratio (CNR),
to assess the performance of double- and triple-intensity correlation-imaging setups. 
The source is a random partially polarized
beam of light obeying Gaussian statistics and the image quality is
evaluated as a function of the degree of polarization (DoP). We show
that the visibility improves when the DoP and the order of imaging increase,
while the SNR behaves oppositely. The CNR is for the most part independent
of the DoP and the imaging order.
The results are important
for the development of new imaging devices using partially polarized
light.
\end{abstract}

\begin{multicols}{2}

\section{Introduction}
\vspace{-2ex}
Ghost imaging, or correlation imaging, is a novel imaging technique in which information about
the object is found  through photon coincidences or classical light correlations
\cite{GATTI08,Erkmen10}. Originally demonstrated with entangled
photon pairs providing a high-visibility image \cite{Pittman95,Strekalov95},
ghost imaging has since been realized with classically correlated \cite{Bennink02} and incoherent (quasi-thermal)
 light \cite{Gatti04a,Valencia05}.  Recently it has been extended
to imaging of temporal objects \cite{Torres08,Setala10,Shirai10}, of pure
phase objects \cite{Shirai11b}, and through aberrations \cite{Cheng09,Wang10,Zhang10,Chan11,Shirai12a}.
Essential features of quantum ghost imaging can be emulated with
classical light,
apart from the visibility as there is a constant background
present in classical imaging \cite{Gatti04b,Gatti04a,Valencia05,Ferri05,Gatti06}.
The advantages of classical light are higher brightness and readily
available sources.  Various techniques have been introduced to improve
the image quality in ghost imaging. These include use of higher-order
correlations \cite{Cao08,Li08}, background subtraction \cite{Chan10},
 differential ghost imaging \cite{Ferri10}, and computational ghost
imaging \cite{Shapiro08,Bromberg09}.

The signal in ghost imaging is noisy and several parameters have been
used to characterize the image quality. Two definitions are typically
employed for visibility \cite{Gatti04a,Cao08}. Besides visibility,
the signal-to-noise ratio (SNR) and the contrast-to-noise ratio (CNR)
characterize image quality and various quantitative measures have been proposed
in ghost imaging for both of them \cite{Agafonov09,Chan10,Ferri10,Brida11}.
While the visibility characterizes the contrast of the image, the SNR and
CNR attempt to take into account the levels of the image noise.
Most of the works concerning image quality in ghost or correlation imaging
have dealt with scalar radiation
and only recently the vectorial nature of light has been considered
by studying the effect that partial polarization has on ghost imaging
with classical light \cite{Liu10,Tong10,Shirai11a}. In particular,
it has been shown that the visibility increases monotonically with
the degree of polarization (DoP) \cite{Shirai11a} and with the order of
intensity correlations \cite{Liu10}, i.e.,
the number of arms in the ghost-imaging setup.

In this work we investigate second-order (double-intensity, two arms)
and third-order (triple-intensity, three arms) classical ghost imaging with partially
polarized light obeying Gaussian statistics. Specifically, we analyze
the image quality in terms of the visibility, SNR, and CNR.  Both
a general intensity-correlation arrangement and a specific ghost-imaging
setup with lenses and an object in one arm are considered.
We show that although the visibility improves with increasing DoP
and is also better for third-order ghost imaging, the situation
is not so straightforward with the SNR and CNR. On the contrary, due
to increased noise, the SNR decreases with the DoP
and the order of imaging, while the CNR is largely not affected by
either. Owing to the noisy source, the SNR and CNR are generally small,
even below one.

The paper is organized as follows. In Sec.~\ref{sec:Foundations-and-geometries}
we discuss electromagnetic intensity correlations and introduce the
double- and triple-intensity ghost-imaging arrangements that we consider. In Sec.~\ref{sec:Visibility}
we analyse the visibility in these setups and in Secs.~\ref{sec:SNR}
and \ref{sec:CNR} we calculate the corresponding SNR and CNR, respectively. A summary
and a brief discussion of our results and their consequences are presented
in Sec.~\ref{sec:Summary-and-conclusion}.
Some mathematical details are relegated
to Appendices \ref{sec:Traces-of}--\ref{sec:Fourth--and-higher-order}.

\section{Electromagnetic correlation imaging  \label{sec:Foundations-and-geometries}}
\vspace{-2ex}

We begin by considering intensity correlation functions (ICFs) of
$N$ random electromagnetic intensities corresponding to fields obeying
joint  Gaussian statistics. We then describe the second-order (double-intensity)
and third-order (triple-intensity) correlation-imaging setups. In both cases
we first introduce a general intensity-correlation arrangement and
then particularize it to ghost imaging with one object arm.

\subsection{Degree of polarization}
\vspace{-1ex}

Consider a stationary, uniformly polarized random beam of light propagating
in the direction of the $z$-axis with the electric field lying in
the $xy$-plane.  The beam's polarization state is characterized
by the polarization matrix \cite{Mandel-95}
\begin{equation}
\mathbf{J}_{0}
=\left[\begin{array}{cc}
J_{xx} & J_{xy}\\
J_{yx} & J_{yy}
\end{array}\right],\label{eq:jii1}
\end{equation}
 where $J_{ij}=\left\langle E_{i}^{*}(t)E_{j}(t)\right\rangle $,
$i,j\in\left\{ x,y\right\} $, is the correlation function between
the electric-field components  and $\left\langle \ldots\right\rangle $
denotes the time (or ensemble) average. Due to the Hermiticity  and
non-negative definiteness of $\mathbf{J}_{0}$ there always exists
a unitary transformation which diagonalizes $\mathbf{J}_{0}$ and
the eigenvalues $J_{1}$ and $J_{2}$ are non-negative. Choosing $J_{1}\geq J_{2}$
the degree of polarization (DoP) is
\begin{equation}
P=\frac{J_{1}-J_{2}}{J_{1}+J_{2}}.\label{eq:p1}
\end{equation}
This expression is mathematically identical to the ones
obtained by dividing the beam into fully polarized and unpolarized parts 
and 
defining $P$ as the ratio between the energy 
of the fully polarized component and that of the total beam \cite{Mandel-95}.
Thus, when the light is completely polarized $P=1$, unpolarized light has $P=0$,
and all other states are partially polarized.

\subsection{Intensity correlations}
\vspace{-1ex}

The equal-time intensity correlation function (ICF) between
$N$ optical intensities is given by \cite{Goodman85,Mandel-95}
\begin{equation}
G^{(N)}(\mathbf{r}_{1},\ldots,\mathbf{r}_{N})=\left\langle I_{1}\cdots I_{N}\right\rangle ,\label{eq:nicf0e}
\end{equation}
where $I_{\alpha}=\mathbf{E}_{\alpha}^{*}(\mathbf{r}_{\alpha},t)\cdot\mathbf{E}_{\alpha}(\mathbf{r}_{\alpha},t)$
with $\alpha\in\{1,\ldots,N\}$ is the instantaneous intensity at point $\mathbf{r}_{\alpha}$. For
stationary light the equal-time ICFs are independent of time but depend on position. The
normalized version of $G^{(N)}$ is
\begin{equation}
g^{(N)}\equiv\frac{\left\langle I_{1}\cdots I_{N}\right\rangle }{\left\langle I_{1}\right\rangle \cdots\left\langle I_{N}\right\rangle },\label{eq:gn1}
\end{equation}
which will be of convenience later. Each intensity can be expressed
as $I_{\alpha}=I_{\alpha x}+I_{\alpha y}=\left|E_{\alpha,x}(\mathbf{r}_{\alpha},t)\right|^{2}+\left|E_{\alpha,y}(\mathbf{r}_{\alpha},t)\right|^{2}$,
where $E_{\alpha,x}(\mathbf{r}_{\alpha},t)$ and $E_{\alpha,y}(\mathbf{r}_{\alpha},t)$
are, respectively, the $x$- and $y$-components of the electric field.
Thus the $N$th-order intensity correlation function can be divided
into a sum of the $N$th-order ICFs of the different components as
\cite{Liu10}
\begin{equation}
\left\langle I_{1}\cdots I_{N}\right\rangle =\sum_{i_{1},\ldots,i_{N}\in\{x,y\}}\left\langle I_{1i_{1}}\cdots I_{Ni_{N}}\right\rangle ,\label{eq:nicf1e}
\end{equation}
where the sum is taken over $i_{\alpha}\in\{x,y\}$ for all $\alpha\in\{1,\ldots,N\}$.
Due to the underlying Gaussian statistics, we may apply the  moment
theorem and obtain \cite{Mandel-95,Cao08}
\begin{equation}
\left\langle I_{1i_{1}}\cdots I_{Ni_{N}}\right\rangle =\sum_{N!}\Gamma_{i_{1}\underline{i_{1}}}^{1\underline{1}}\cdots\Gamma_{i_{N}\underline{i_{N}}}^{N\underline{N}},\label{eq:nicf1s}
\end{equation}
where the summation is performed over all the $N!$ possible permutations
of the underlined indices and $\Gamma_{ij}^{\alpha\beta}\equiv\bigl\langle E_{\alpha,i}^{*}(\mathbf{r}_{\alpha},t)E_{\beta,j}(\mathbf{r}_{\beta},t)\bigr\rangle,$
with $i,j\in\{x,y\}$ and $\alpha,\beta\in\{1,\ldots,N\}$, is the
correlation function between the electric-field components.

We can express the $N$th-order ICF in terms of the equal-time mutual coherence (or the mutual intensity) matrix
\begin{equation}
\mathbf{\Gamma}^{\alpha\beta}=\left[\begin{array}{cc}
\Gamma_{xx}^{\alpha\beta} & \Gamma_{xy}^{\alpha\beta}\\
\Gamma_{yx}^{\alpha\beta} & \Gamma_{yy}^{\alpha\beta}
\end{array}\right],\label{eq:mcfm1}
\end{equation}
where the elements are defined as above. In the second-order we
find that
\begin{equation}
\left\langle I_{1}I_{2}\right\rangle =\mathrm{tr}\mathbf{\Gamma}^{11}\mathrm{tr}\mathbf{\Gamma}^{22}+\mathrm{tr}(\mathbf{\Gamma}^{12}\mathbf{\Gamma}^{21}),\label{eq:2icf1e}
\end{equation}
where $\mathrm{tr}$ denotes the trace and $\mathrm{tr}\mathbf{\Gamma}^{\alpha\alpha}=\left\langle I_{\alpha}\right\rangle $
is the $\alpha$th average intensity.  The first term is the product
of the mean intensities $\left\langle I_{1}\right\rangle \left\langle I_{2}\right\rangle $
and the second one is the correlation of the intensity fluctuations
$\left\langle \Delta I_{1}\Delta I_{2}\right\rangle $, where $\Delta I_{\alpha}=I_{\alpha}-\left\langle I_{\alpha}\right\rangle$
with $\alpha\in\{1,2\}$. It is the latter term that is responsible for image
formation in ghost imaging. Similarly, the third-order ICF takes
on the form
\begin{alignat}{1}
\left\langle I_{1}I_{2}I_{3}\right\rangle  & =\mathrm{tr}\mathbf{\Gamma}^{11}\mathrm{tr}\mathbf{\Gamma}^{22}\mathrm{tr}\mathbf{\Gamma}^{33}+\mathrm{tr}\mathbf{\Gamma}^{33}\mathrm{tr}(\mathbf{\Gamma}^{12}\mathbf{\Gamma}^{21})\nonumber \\
 & +\mathrm{tr}\mathbf{\Gamma}^{22}\mathrm{tr}(\mathbf{\Gamma}^{13}\mathbf{\Gamma}^{31})+\mathrm{tr}\mathbf{\Gamma}^{11}\mathrm{tr}(\mathbf{\Gamma}^{23}\mathbf{\Gamma}^{32})\nonumber \\
 & +\mathrm{tr}(\mathbf{\Gamma}^{12}\mathbf{\Gamma}^{23}\mathbf{\Gamma}^{31})+\mathrm{tr}(\mathbf{\Gamma}^{13}\mathbf{\Gamma}^{32}\mathbf{\Gamma}^{21}).\label{eq:3icf1e}
\end{alignat}
The first term is again the product of the mean intensities $\left\langle I_{1}\right\rangle \left\langle I_{2}\right\rangle \left\langle I_{3}\right\rangle $,
the next three terms have a mean intensity multiplied by the correlation
of intensity fluctuations between the remaining two intensities, e.g.,
$\left\langle I_{3}\right\rangle \left\langle \Delta I_{1}\Delta I_{2}\right\rangle $,
and the sum of the last two terms is equal to the correlation of all
the intensity fluctuations $\left\langle \Delta I_{1}\Delta I_{2}\Delta I_{3}\right\rangle $.

\subsection{Double-intensity correlation imaging\label{sub:Double-intensity}}
\vspace{-1ex}

\begin{figure}[H]
\centering
\includegraphics[width=84mm]{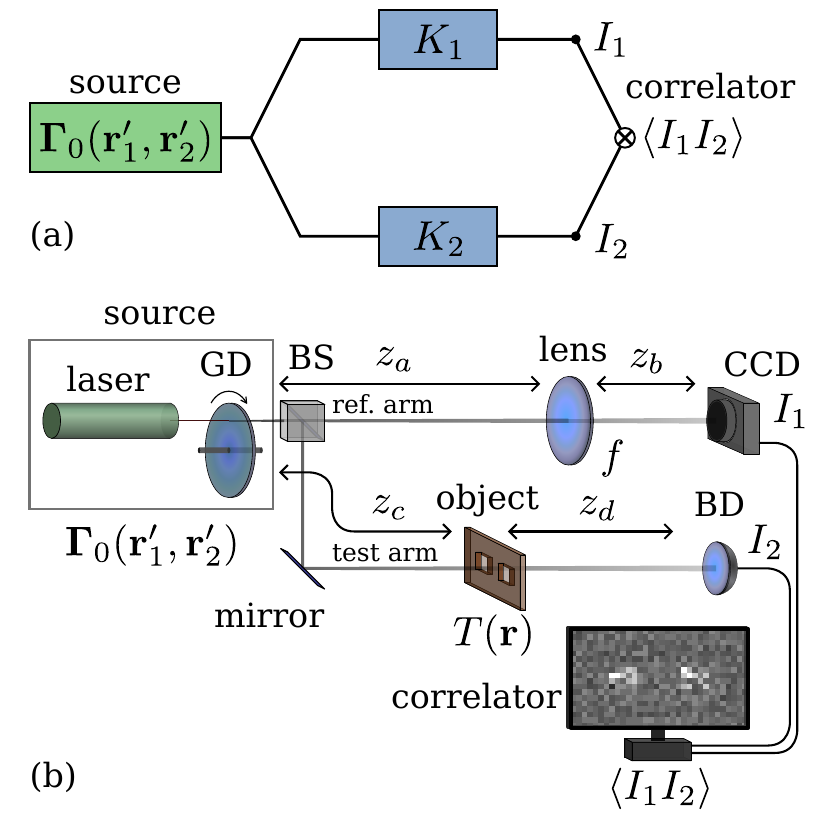}
\caption{\small (a) Generic and (b) specific double-intensity correlation-imaging setups.
In (a), the source beam [$\mathbf{\Gamma}_{0}(\mathbf{r}_{1}^{\prime},\mathbf{r}_{2}^{\prime})$]
is split into two arms characterized by the kernels $K_{1}$ and $K_{2}$.
The beam intensities $I_{1}$ and $I_{2}$ at the end of the arms
are measured and correlated ($\left\langle I_{1}I_{2}\right\rangle $).
In (b), a laser and a rotating ground glass
disk (GD) are used to create a spatially incoherent light beam which then is
divided into two arms with a beam splitter (BS). The reference
arm has a lens with the focal length $f$ and the test arm contains
the object with the transmission function $T(\mathbf{r})$. The propagation distances
are denoted by $z_{i}$, $i\in\{a,b,c,d\}$.
The reference arm has a CCD camera to measure the intensity distribution
and the test arm includes a bucket detector (BD) which measures the total
intensity. An intensity correlation ($\left\langle I_{1}I_{2}\right\rangle $) is performed to form an image
of the object. \label{fig:double}}
\end{figure}

A general second-order correlation imaging geometry is depicted in
Fig.~\ref{fig:double}(a). The source is a classical, quasi-monochromatic, uniformly
partially polarized electromagnetic beam of light represented by the
mutual intensity matrix $\mathbf{\Gamma}_{0}(\mathbf{r}_{1}^{\prime},\mathbf{r}_{2}^{\prime})=\mathbf{J}_{0}\gamma_{0}(\mathbf{r}_{1}^{\prime},\mathbf{r}_{2}^{\prime})$,
where $\mathbf{J}_{0}$ is the polarization matrix and $\gamma_{0}(\mathbf{r}_{1}^{\prime},\mathbf{r}_{2}^{\prime})$
is a normalized spatial coherence function. The source beam is split
into two arms and propagated along the paths described by the kernels
$K_{1}$ and $K_{2}$. The arms are taken to be polarization independent and their
properties do not vary with time. The intensities $I_{1}$ and $I_{2}$
at the end of the paths are measured as a function of position and
correlated producing $\left\langle I_{1}I_{2}\right\rangle $.

The intensity correlation is given by Eq.~\eqref{eq:2icf1e}. The
mutual intensity matrices appearing in this equation can be written
as  \cite{Tong10,Shirai11a}
\begin{equation}
\mathbf{\Gamma}^{\alpha\beta}=\mathbf{J}_{0}\hat{\Gamma}^{\alpha\beta},\label{eq:Gab2r}
\end{equation}
where
\begin{equation}
\hat{\Gamma}^{\alpha\beta}=\iint\gamma_{0}(\mathbf{r}_{1}^{\prime},\mathbf{r}_{2}^{\prime})K_{\alpha}^{*}(\mathbf{r}_{\alpha},
\mathbf{r}_{1}^{\prime})K_{\beta}(\mathbf{r}_{\beta},\mathbf{r}_{2}^{\prime})\mathrm{d}\mathbf{r}_{1}^{\prime}
\mathrm{d}\mathbf{r}_{2}^{\prime}\label{eq:Gab3r}
\end{equation}
with $\alpha,\beta\in\left\{ 1,2\right\} $. Here and henceforth the
primed coordinates refer to the source plane and the coordinates $\mathbf{r}_{\alpha}$
refer to the plane of the detector in arm $\alpha$. We define
the normalized form of $\hat{\Gamma}^{\alpha\beta}$ as
\begin{equation}
\hat{\gamma}^{\alpha\beta}\equiv\hat{\Gamma}^{\alpha\beta}/\sqrt{\hat{\Gamma}^{\alpha\alpha}\hat{\Gamma}^{\beta\beta}},
\label{eq:gab3r}
\end{equation}
which in view of the Schwarz inequality satisfies $0\leq|\hat{\gamma}^{\alpha\beta}|\leq1$.
Inserting Eq.~\eqref{eq:Gab2r} into Eq.~\eqref{eq:2icf1e} and employing Eqs.~\eqref{eq:gn1} and \eqref{eq:gab3r} gives
the normalized intensity correlation function
\begin{equation}
g^{(2)}(\mathbf{r}_{1},\mathbf{r}_{2})=1+\frac{P^{2}+1}{2}\left|\hat{\gamma}^{12}\right|^{2},\label{eq:2icf2e}
\end{equation}
where we used Eq.~\eqref{eq:trndiv} of Appendix \ref{sec:Traces-of}.
The quantity $g^{(2)}$ has a constant background and a term that
contains information on the correlation of the intensities $I_{1}$
and $I_{2}$ and depends on the degree of polarization $P$.

\paragraph{Double-intensity ghost-imaging setup}

We now consider the specific ghost-imaging setup shown in Fig.~\ref{fig:double}(b).
The source is spatially completely incoherent, characterized by the mutual intensity matrix $\mathbf{\Gamma}_{0}(\mathbf{r}_{1}^{\prime},\mathbf{r}_{2}^{\prime})
=\mathbf{J}_{0}\delta(\mathbf{r}_{2}^{\prime}-\mathbf{r}_{1}^{\prime})$,
where $\delta$ is the Dirac delta function. In addition, the reference
arm ($K_{1}$) contains a lens and a high-resolution detector,
and the test arm ($K_{2}$) includes the object and a bucket detector
with no spatial resolution.

For a thin and large-aperture lens, the reference arm is described by the
kernel \cite{Goodman96}
\begin{alignat}{1}
 K_1(\mathbf{r}_1,\mathbf{r}_1^{\prime}) & =\frac{i}{\lambda\zeta}\exp\left\{ \frac{ik}{2\zeta}\left[\left(\mathbf{r}_1^{\prime}-\mathbf{r}_1\right)^{2}
 -\frac{z_{b}{\mathbf{r}_1^{\prime}}^{2}+z_{a}\mathbf{r}_1^{2}}{f}\right]\right\} ,\label{eq:klens1}
\end{alignat}
where $z_{a}$ and $z_{b}$ are the propagation distances, $f$
is the lens focal length, $\zeta\equiv z_{a}+z_{b}-z_{a}z_{b}/f$,
and $k=2\pi/\lambda$ is the wave number, with $\lambda$ being the
wavelength. The transmission function $T(\mathbf{r})$ characterizes
the object, which is preceded with the propagation distance $z_{c}$
and succeeded with $z_{d}$. The kernel of the test arm is
\begin{alignat}{1}
 & K_2(\mathbf{r}_2,\mathbf{r}_2^{\prime})\nonumber \\
 & =\frac{1}{\lambda^{2}z_{a}z_{b}}\int T(\mathbf{r})\exp\left\{ \frac{ik}{2}\left[\frac{\left(\mathbf{r}-\mathbf{r}_2^{\prime}\right)^{2}}{z_{c}}
 +\frac{\left(\mathbf{r}_2-\mathbf{r}\right)^{2}}{z_{d}}\right]\right\} \mathrm{d}\mathbf{r}.\label{eq:ktest1}
\end{alignat}
The normalized correlation between the output intensities is given
by Eq.~\eqref{eq:2icf2e}, where $\hat{\gamma}^{12}$ is specified
by {[}see Eqs.~\eqref{eq:Gab3r} and \eqref{eq:gab3r}{]}
\begin{equation}
\hat{\Gamma}^{\alpha\beta}=\int K_{\alpha}^{*}(\mathbf{r}_{\alpha},\mathbf{r}^{\prime})K_{\beta}(\mathbf{r}_{\beta},\mathbf{r}^{\prime})\mathrm{d}\mathbf{r}^{\prime}\label{eq:Gab3rdelta}
\end{equation}
with $\alpha,\beta\in\{1,2\}$. If the imaging condition \cite{Cao05,Cai05b}
\begin{equation}
\frac{1}{f}=\frac{1}{z_{a}-z_{c}}+\frac{1}{z_{b}} \label{eq:image1}
\end{equation}
holds, we find with the use of Eqs.~\eqref{eq:klens1}--\eqref{eq:image1} that
\begin{equation}
\left|\hat{\Gamma}^{12}\right|^{2}=\frac{\left(z_{a}-z_{c}\right)^{2}}{\lambda^{2}z_{b}^{2}z_{d}^{2}}\left|T(-\frac{z_{a}-z_{c}}{z_{b}}\mathbf{r}_{1})\right|^{2}.\label{eq:Ga33rdelta}
\end{equation}
Thus an intensity correlation measurement can form an image (of the absolute value) of the
object. The image is not due to the output intensities {[}$\mathrm{tr}\mathbf{\Gamma}^{11}$,
$\mathrm{tr}\mathbf{\Gamma}^{22}$ in Eq.~\eqref{eq:2icf1e}{]}
but rather results from the correlation of the intensity fluctuations
at the outputs {[}$\mathrm{tr}(\mathbf{\Gamma}^{12}\mathbf{\Gamma}^{21})${]}.
Since $0\leq|\hat{\gamma}^{12}|\leq1$, the normalized output intensity
correlation is in the range $1\leq g^{(2)}\leq\left(P^{2}+3\right)/2$
 suggesting that the image quality depends on the DoP. The range
of $g^{(2)}$ holds both for the general geometry of Fig.~\ref{fig:double}(a)
and for the specific setup shown in Fig.~\ref{fig:double}(b)
indicating that they have the same image quality.

\subsection{Triple-intensity correlation imaging\label{sub:Triple-intensity}}
\vspace{-1ex}

\begin{figure}[H]
\centering
\includegraphics[width=84mm]{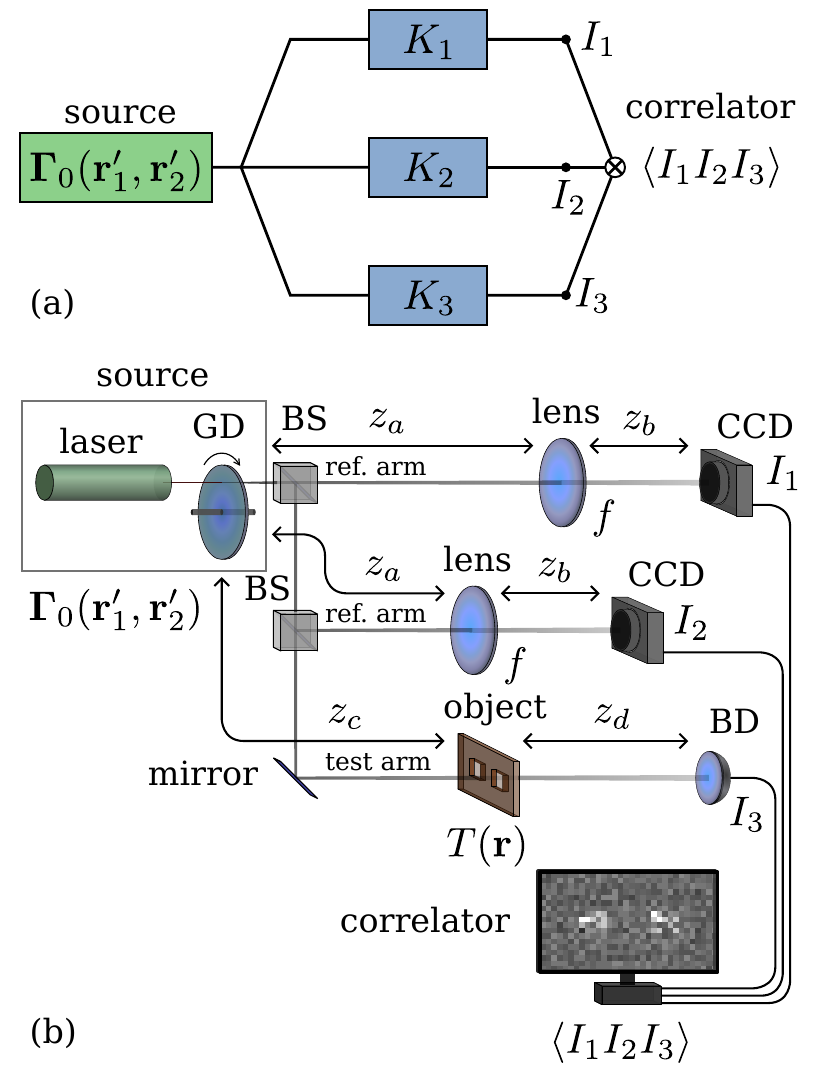}
\caption{\small (a) Generic and (b) specific triple-intensity correlation
imaging setups. Part (a) is similar to Fig.~\ref{fig:double}(a)
but has three arms instead of two. Likewise, part (b) is analogous to
Fig.~\ref{fig:double}(b) with the exception that there is an additional
reference arm which is identical to the first one. \label{fig:triple}
}
\end{figure}

A general third-order correlation imaging geometry containing three
arms described by the kernels $K_{1}$, $K_{2}$, and $K_{3}$ is
depicted in Fig.~\ref{fig:triple}(a). We make the same assumptions
about the source ($\mathbf{\Gamma}_{0}$) and the arms as in the second-order
case analyzed in the beginning of Sec.~\ref{sub:Double-intensity}.
The correlation of the intensities at the end of the arms, $\left\langle I_{1}I_{2}I_{3}\right\rangle $,
is provided by Eq.~\eqref{eq:3icf1e}. The mutual intensity matrices
appearing in this equation satisfy Eqs.~\eqref{eq:Gab2r} and \eqref{eq:Gab3r}
with $\alpha,\beta\in\{1,2,3\}$. These equations together with Eqs.~\eqref{eq:gn1},
\eqref{eq:3icf1e}, \eqref{eq:gab3r}, and \eqref{eq:trndiv} give
the normalized triple-intensity ICF
\begin{alignat}{1}
g^{(3)}(\mathbf{r}_{1},\mathbf{r}_{2},\mathbf{r}_{3}) & =1+\frac{P^{2}+1}{2}(\left|\hat{\gamma}^{12}\right|^{2}+\left|\hat{\gamma}^{13}\right|^{2}
+\left|\hat{\gamma}^{23}\right|^{2})\nonumber \\
 & +\frac{3P^{2}+1}{2}\Re(\hat{\gamma}^{12}\hat{\gamma}^{23}\hat{\gamma}^{31}),\label{eq:3icf2e1}
\end{alignat}
where $\Re(\ldots)$ denotes the real part. We can express the
complex correlation coefficients in terms of their magnitudes and phases
as $\hat{\gamma}^{\alpha\beta}=|\hat{\gamma}^{\alpha\beta}|e^{i\phi_{\alpha\beta}}$. This implies that $\Re(\hat{\gamma}^{12}\hat{\gamma}^{23}\hat{\gamma}^{31})
=|\hat{\gamma}^{12}||\hat{\gamma}^{23}||\hat{\gamma}^{13}|\cos(\Delta\phi_{123})$,
where we have introduced the notation $\Delta\phi_{123}\equiv\phi_{12}+\phi_{23}-\phi_{13}$
and used the fact that $(\hat{\gamma}^{\alpha\beta})^*=\hat{\gamma}^{\beta\alpha}$.
The function $g^{(3)}$ has a constant background (equal to one)
followed by three DoP-dependent terms which each contain information
on the correlation of a different pair of fields \cite{Li08}. The last
term has the strongest dependence on $P$ and it carries information
on all the three correlations. Because $0\leq|\hat{\gamma}^{\alpha\beta}|\leq1$
for all $\alpha,\beta\in\{1,2,3\}$, the normalized third-order ICF
is in the range $1\leq g^{(3)}\leq3P^{2}+3$ suggesting, as in the second-order
case, that the image quality depends in an essectial way on the DoP of the
source.

\paragraph{Triple-intensity setup with one object arm}

As the last arrangement, we consider the specific third-order ghost-imaging scheme
shown in Fig.~\ref{fig:triple}(b). The source ($\mathbf{\Gamma}_{0}$),
the reference arms {[}$K_{1}$ and $K_{2}$, as expressed in Eq.~\eqref{eq:klens1}{]},
and the test arm {[}$K_{3}$, see Eq.~\eqref{eq:ktest1}{]} are as
in the specific double-intensity ghost-imaging setup of Sec.~\ref{sub:Double-intensity}.
The normalized ICF of the outputs is given by Eq.~\eqref{eq:3icf2e1},
where the functions $\hat{\gamma}^{\alpha\beta}$ are obtained from
Eqs.~\eqref{eq:gab3r} and \eqref{eq:Gab3rdelta} with $\alpha,\beta\in\{1,2,3\}$.

Since both reference arms are the same, the correlation function between them is
\begin{alignat}{1}
\hat{\Gamma}^{12} & =\exp\left\{ \frac{ik}{2}\left[\frac{1-z_{a}/f}{\zeta}\left(\mathbf{r}_{2}^{2}-\mathbf{r}_{1}^{2}\right)\right]\right\} \delta\left(\mathbf{r}_{1}-\mathbf{r}_{2}\right),\label{eq:G123rdelta}
\end{alignat}
where the parameters are the same as those in Eq.~\eqref{eq:klens1}. The Dirac
delta function in this equation is due to the spatially fully incoherent
source. Physically $\hat{\Gamma}^{12}$ in Eq.~\eqref{eq:G123rdelta} represents
complete lack of correlation between the two fields when $\mathbf{r}_1 \neq \mathbf{r}_2$
but shows complete correlation (with a uniform amplitude of unity) when the points are equal.
In Appendix~\ref{sec:Partially-coherent-source} we use a spatially partially
coherent source to explicitly demonstrate that the normalized correlation
function indeed is \foreignlanguage{british}{$\hat{\gamma}^{12}=1$} when
$\mathbf{r}_{1}=\mathbf{r}_{2}$. The remaining correlation quantities
$\hat{\Gamma}^{13}$ and $\hat{\Gamma}^{23}$ are equal and given
by Eq.~\eqref{eq:Ga33rdelta} as long as the imaging condition of Eq.~\eqref{eq:image1}
holds. For $\mathbf{r}_{1}=\mathbf{r}_{2}$ the triple-intensity ICF of Eq.~\eqref{eq:3icf2e1} then yields
\begin{equation}
g^{(3)}(\mathbf{r}_{1},\mathbf{r}_{1},\mathbf{r}_{3})=\frac{P^{2}+3}{2}
+\frac{5P^{2}+3}{2}\left|\hat{\gamma}^{13}\right|^{2}.\label{eq:3icf2e2}
\end{equation}
In other words, the information on the object is in the terms
$\mathrm{tr}(\mathbf{\Gamma}^{13}\mathbf{\Gamma}^{31})$, $\mathrm{tr}(\mathbf{\Gamma}^{23}\mathbf{\Gamma}^{32})$,
$\mathrm{tr}(\mathbf{\Gamma}^{12}\mathbf{\Gamma}^{23}\mathbf{\Gamma}^{31})$,
and $\mathrm{tr}(\mathbf{\Gamma}^{13}\mathbf{\Gamma}^{32}\mathbf{\Gamma}^{21})$
of Eq.~\eqref{eq:3icf1e}, characterizing the correlations of the
intensity fluctuations between the test arm and the reference arms.
Because now $\hat{\gamma}^{12}=1$ and $0\leq|\hat{\gamma}^{13}|\leq1$,
the normalized third-order intensity correlation of Eq.~\eqref{eq:3icf2e2} is in the range
$\left(P^{2}+3\right)/2\leq g^{(3)}\leq3P^{2}+3$.

\section{Visibility\label{sec:Visibility}}
\vspace{-2ex}

Among the ghost-imaging quality parameters, the visibility of the image has
remained as an important quantity, although other
parameters have been introduced. Visibility (or image contrast) describes
the relative difference between the bright and dark areas of the image.
Most studies on visibility in ghost imaging were initially performed in the double-intensity case
\cite{Gatti04a,Gatti06,GATTI08}, but higher-order imaging has
gained increased attention \cite{Cao08,Agafonov09,Chan09}.
Some preliminary works on the influence of the degree of polarization
on visibility have been carried out recently \cite{Liu10,Shirai11a}.

\begin{figure}[H]
\centering\includegraphics[width=84mm]{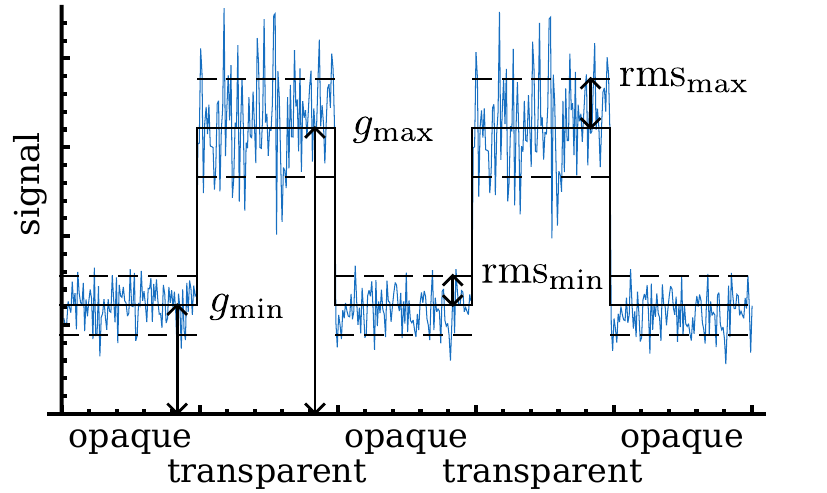}
\caption{\small Qualitative sample realization of a normalized ghost-imaging signal.
The object has two transparent regions with a normalized mean signal $g_{\mathrm{max}}$ corresponding to
the bright areas of the image. The rest of the object is opaque leading
to the signal $g_{\mathrm{min}}$ and the dark areas in the image. The
noise levels in the bright and dark areas are characterized by the root-mean-square
values $\mathrm{rms}_{\mathrm{max}}$ and $\mathrm{rms}_{\mathrm{min}}$ of the fluctuations,
respectively. \label{fig:gaussian}}
\end{figure}

Two main definitions are in use for the visibility in ghost imaging.
The definition employed by Cao \emph{et al.} \cite{Cao08} for $N$th-order imaging is
\begin{equation}
V_{\mathrm{C}}^{(N)}\equiv\frac{g_{\mathrm{max}}^{(N)}-g_{\mathrm{min}}^{(N)}}{g_{\mathrm{max}}^{(N)}
+g_{\mathrm{min}}^{(N)}},\label{eq:visnc01}
\end{equation}
where $g_{\mathrm{max}}^{(N)}$ and $g_{\mathrm{min}}^{(N)}$ can be thought of as the average
signal in the bright and dark areas of the ghost image, corresponding to the
parts where the object is transparent or opaque, respectively. This idea is illustrated in Fig.~\ref{fig:gaussian},
where a one-dimensional random realization of a signal related to
an object with fully transparent and opaque regions is shown.

The visibility employed by Gatti \emph{et al.} \cite{Gatti06,GATTI08} may be
defined for second-order ghost imaging as $V_{\mathrm{G}}^{(2)}\equiv\left\langle \Delta I_{1}\Delta I_{2}\right\rangle _{\mathrm{max}}/\left\langle I_{1}I_{2}\right\rangle _{\mathrm{max}}$.
Since $\left\langle I_{1}\right\rangle \left\langle I_{2}\right\rangle$
is a constant background, we obtain
\begin{equation}
V_{\mathrm{G}}^{(2)}=\frac{\left\langle I_{1}I_{2}\right\rangle _{\mathrm{max}}-\left\langle I_{1}\right\rangle \left\langle I_{2}\right\rangle }{\left\langle I_{1}I_{2}\right\rangle _{\mathrm{max}}}=\frac{g_{\mathrm{max}}^{(2)}-1}{g_{\mathrm{max}}^{(2)}}.\label{eq:vis2g01-1}
\end{equation}
There are two principal ways of extending this to higher orders,
one of which assumes that
$\left\langle I_{1}\cdots I_{N}\right\rangle _{\mathrm{min}}=\left\langle I_{1}\right\rangle \cdots\left\langle I_{N}\right\rangle $
\cite{LiuArxiv} and the other that $\left\langle I_{1}\cdots I_{N}\right\rangle _{\mathrm{min}}=\left\langle I_{1}\cdots I_{N-1}\right\rangle \left\langle I_{N}\right\rangle $
\cite{Agafonov09}. Translated to the triple-intensity imaging schemes
of Fig.~\ref{fig:triple}, the first case corresponds to part (a); the minimum of the ghost-imaging signal is
obtained when all intensity-fluctuation correlations disappear
and only the first term in Eq.~\eqref{eq:3icf1e} remains.
The second case, $\left\langle I_{1}I_{2}I_{3}\right\rangle_{\mathrm{min}}=\left\langle I_{1}I_{2}\right\rangle \left\langle I_{3}\right\rangle $, is closely related to Fig.~\ref{fig:triple}(b), where the intensity correlation
between the reference arms {[}the first two terms in Eq.~\eqref{eq:3icf1e}{]}
contributes to the background, since these beams contain no information about the image.
However, we want to take into account all the various arrangements
and thus we generalize the definition to the $N$th order as
\begin{equation}
V_{\mathrm{G}}^{(N)}\equiv\frac{\left\langle I_{1}\cdots I_{N}\right\rangle _{\mathrm{max}}-\left\langle I_{1}\cdots I_{N}\right\rangle _{\mathrm{min}}}{\left\langle I_{1}\cdots I_{N}\right\rangle _{\mathrm{max}}}=\frac{g_{\mathrm{max}}^{(N)}-g_{\mathrm{min}}^{(N)}}{g_{\mathrm{max}}^{(N)}}.\label{eq:visng01}
\end{equation}
The visibilities are normalized so that $0\leq V_{C}^{(N)},V_{G}^{(N)}\leq1$,
but they scale differently between the end points. However, both definitions, Eqs.~\eqref{eq:visnc01}
and \eqref{eq:visng01}, can be expected to lead to similar
physical conclusions on the image quality. Next we make use of the results
from Sec.~\ref{sec:Foundations-and-geometries} to calculate these
quantities for double- and triple-intensity ghost imaging as a function
of the degree of polarization.

\subsection{Double-intensity correlation imaging}
\vspace{-1ex}

The correlation-imaging setups of Fig.~\ref{fig:double} can both
be described by the normalized second-order ICF of Eq.~\eqref{eq:2icf2e}.
Using the facts that $|\hat{\gamma}^{12}|_{\mathrm{min}}=0$ and $|\hat{\gamma}^{12}|_{\mathrm{max}}=1$,
the visibilities according to Eqs.~\eqref{eq:visnc01}
and \eqref{eq:visng01} then are \cite{Shirai11a}
\begin{alignat}{1}
V_{\mathrm{C}}^{(2)} & =\frac{P^{2}+1}{P^{2}+5},\label{eq:vis2c02e}\\
V_{\mathrm{G}}^{(2)} & =\frac{P^{2}+1}{P^{2}+3}.\label{eq:vis2g02e}
\end{alignat}
Equations \eqref{eq:vis2c02e} and \eqref{eq:vis2g02e} correspond to the maximum theoretical
visibilities and they are illustrated with the solid lines in Figs.~\ref{fig:vis}(a) and \ref{fig:vis}(b),
respectively. Both curves indicate that a rise in the degree
of polarization affects the image visibility positively.

\subsection{Triple-intensity correlation imaging\label{sub:Triple-intensity-visibility}}
\vspace{-1ex}

\begin{figure}[H]
\centering
\includegraphics[width=84mm]{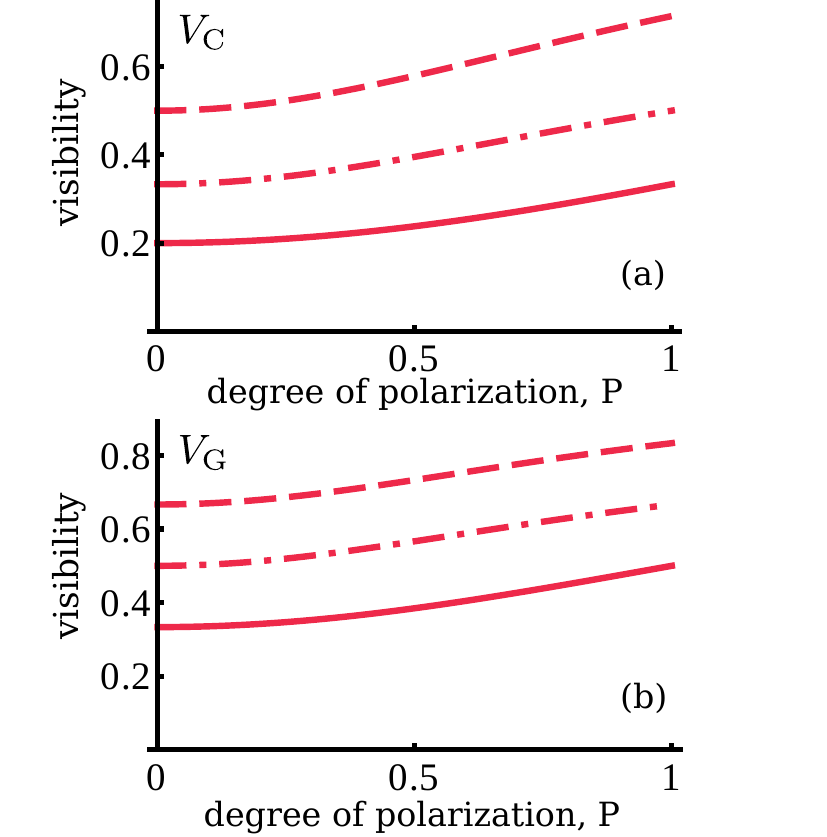}
\caption{\small Comparison of the maximum visibility according
to the definition by (a) Cao \emph{et al.} [Eq.~\eqref{eq:visnc01}]
and (b) Gatti \emph{et al.} [Eq.~\eqref{eq:visng01}].
The solid lines correspond to the double-intensity case [Eqs.~\eqref{eq:vis2c02e}
and \eqref{eq:vis2g02e}], the dashed lines [Eqs.~\eqref{eq:vis3c02e}
and \eqref{eq:vis3g02e}] and dash-dotted lines [Eqs.~\eqref{eq:vis3c03e}
and \eqref{eq:vis3g03e}] are for the general and specific
triple-intensity setups, respectively. \label{fig:vis}}
\end{figure}

The normalized output of a general triple-intensity correlation-imaging
setup is given by Eq.~\eqref{eq:3icf2e1}. Using the values $|\hat{\gamma}^{\alpha\beta}|_{\mathrm{min}}=0$,
$|\hat{\gamma}^{\alpha\beta}|_{\mathrm{max}}=1$ (for $\alpha,\beta\in\{1,2,3\}$)
and $\cos(\Delta\phi_{123})=1$, the visibilities from Eqs.~\eqref{eq:visnc01}
and \eqref{eq:visng01} take on their maximum values and are \cite{Kellock11c}
\begin{alignat}{1}
V_{\mathrm{C}}^{(3)} & =\frac{3P^{2}+2}{3P^{2}+4},\label{eq:vis3c02e}\\
V_{\mathrm{G}}^{(3)} & =\frac{3P^{2}+2}{3P^{2}+3}.\label{eq:vis3g02e}
\end{alignat}
Equation \eqref{eq:vis3c02e} agrees with Eq.~(17) in \cite{Liu10} and both Eqs.~\eqref{eq:vis3c02e}
and \eqref{eq:vis3g02e} lead to the same physical conclusions that
the visibility increases as a function of the DoP, as can be
seen from the dashed lines in Fig.~\ref{fig:vis}. The visibility
in the general triple-intensity imaging scheme is considerably higher
than that of double-intensity imaging, for each value of the degree of
polarization.

\paragraph{Triple-intensity setup with one object arm}

Lastly, we examine the visibility related to the specific arrangement
in Fig.~\ref{fig:triple}(b) which has the normalized output
of Eq.~\eqref{eq:3icf2e2}. As shown in Sec.~\ref{sub:Triple-intensity},
in contrast to the general triple-intensity case, the normalized correlation
between the first two arms in this setup is equal to unity ($|\hat{\gamma}^{12}|=1$)
and it does not contribute to the image formation. For the remaining
correlation parameters we have $|\hat{\gamma}^{13}|=|\hat{\gamma}^{23}|$,
$|\hat{\gamma}^{13}|_{\mathrm{min}}=0$, and $|\hat{\gamma}^{13}|_{\mathrm{max}}=1$.
Using these, the visibilities according to Eqs.~\eqref{eq:visnc01}
and \eqref{eq:visng01} become
\begin{alignat}{1}
V_{\mathrm{C}}^{(3)} & =\frac{5P^{2}+3}{7P^{2}+9},\label{eq:vis3c03e}\\
V_{\mathrm{G}}^{(3)} & =\frac{5P^{2}+3}{6P^{2}+4}.\label{eq:vis3g03e}
\end{alignat}
Equations~\eqref{eq:vis3c03e} and \eqref{eq:vis3g03e} are illustrated
with the dash-dotted lines in Figs.~\ref{fig:vis}(a) and \ref{fig:vis}(b).
The visibility of the specific triple-intensity ghost-imaging setup behaves similarly
to that of the two previous cases we considered above as the DoP is varied. It is smaller
than the visibility in the general triple-intensity
case but larger than in double-intensity ghost imaging, for all $P$.

All the visibilities improve with the increase of the DoP. This is
caused by the fact that while the backgrounds are independent or weakly
dependent on $P$, the correlation terms (which contain
the image) show a stronger polarization dependence, as is evidenced for instance by
Eqs.~\eqref{eq:2icf2e}, \eqref{eq:3icf2e1}, and \eqref{eq:3icf2e2}{]}.
For each $P$, the visibility in a given arrangement
is proportional to the relative abundance of the correlation-term contributions
over the background.

\section{Signal-to-noise ratio\label{sec:SNR}}
\vspace{-2ex}

The visibility is a measure of the (relative) difference between
the light ($g_{\mathrm{max}}^{(N)}$) and dark ($g_{\mathrm{min}}^{(N)}$)
areas of the image. However, it contains no information about the noise, i.e.,
the strength of the intensity fluctuations within the image. We
have assumed the source to exhibit electric field-vector fluctuations obeying Gaussian
statistics. For such a source, $\langle (\Delta I)^2\rangle /\langle I \rangle^2 =
(P^2 + 1)/2$, showing that the higher $P$, the larger are the intensity
fluctuations in relation to the mean value \cite{Mandel-95,Setala04a}. Hence the source noise level grows as $P$
increases. The fluctuations are at their minimum $\langle (\Delta I)^2\rangle = \langle I \rangle^2 /2$ 
when $P = 0$, indicating that the noise is comparable to the signal at the source.
The source fluctuations will produce a certain amount of
noise to the outputs, which we calculate for both double- and triple-intensity
ghost imaging. Other sources of noise, such as those in the
optical system or the detection, are not taken into account.

In $N$th-order ghost imaging, the quantity $I_{1}\cdots I_{N}$ is the fluctuating signal. Its variance is
$\mathrm{var}\left(I_{1}\cdots I_{N}\right)\equiv\bigl\langle\left(I_{1}\cdots I_{N}
-\left\langle I_{1}\cdots I_{N}\right\rangle \right)^{2}\bigr\rangle$
and the related noise is the square
root of the variance (or the standard deviation), i.e.,
\begin{equation}
\mathrm{noise}\left(I_{1}\cdots I_{N}\right)=\sqrt{\left\langle I_{1}^{2}\cdots I_{N}^{2}\right\rangle -\left\langle I_{1}\cdots I_{N}\right\rangle ^{2}}.\label{eq:noise01}
\end{equation}
This is the root-mean-square (rms) of the deviation from the mean
signal and it is always non-zero for Gaussian statistics. We define the signal-to-noise ratio (SNR) as
\begin{equation}
\mathrm{SNR}^{(N)}\equiv\frac{\left\langle I_{1}\cdots I_{N}\right\rangle }{\mathrm{noise}\left(I_{1}\cdots I_{N}\right)},\label{eq:snrn01}
\end{equation}
i.e., the  average signal divided by the associated rms noise.

The transparent parts of the object lead to bright areas in the ghost image
with the normalized average signal $g_{\mathrm{max}}$
and the average noise $\mathrm{rms}_{\mathrm{max}}$, as is illustrated in Fig.~\ref{fig:gaussian}.
The maximum signal is obtained when
the correlation parameters in Eqs.~\eqref{eq:2icf2e}, \eqref{eq:3icf2e1},
and \eqref{eq:3icf2e2} take on their maximum values. The SNR related
to the bright areas is denoted by $\mathrm{SNR}[g_{\mathrm{max}}^{(N)}]$.
On the other hand, the opaque parts of the object lead to dark image areas, and these are characterized by
$g_{\mathrm{min}}$ and $\mathrm{rms}_{\mathrm{min}}$ in Fig.~\ref{fig:gaussian}.
The signals for the dark areas are attained when the correlation
parameters reach their minima, and the related SNR is denoted by $\mathrm{SNR}[g_{\mathrm{min}}^{(N)}]$.
In addition to the SNRs of the bright and dark areas, we examine the minimum and maximum SNRs within
the ghost image, denoted by $\mathrm{SNR}_{\mathrm{min}}^{(N)}$
and $\mathrm{SNR}_{\mathrm{max}}^{(N)}$, respectively.
It can happen, for example, that the SNR maximum is obtained when the signal achieves
its minimum (i.e., $\mathrm{SNR}_{\mathrm{max}}^{(N)}\approx\mathrm{SNR}[g_{\mathrm{min}}^{(N)}]$).

For later use it is convenient, in view of Eq.~\eqref{eq:noise01}, to introduce the notation
\begin{equation}
  \tilde{g}^{(2N)}
  \equiv\left\langle I_{1}^{2}\cdots I_{N}^{2}\right\rangle/\left\langle I_{1}\right\rangle ^{2}\cdots\left\langle I_{N}\right\rangle ^{2} \label{gr2N}
\end{equation}
for the $2N$th-order ICF of $N$ different pairs of intensities. The SNR can be written as
\begin{equation}
  \mathrm{SNR}^{(N)}
  =\frac{g^{(N)}}{\sqrt{\tilde{g}^{(2N)}-\left[g^{(N)}\right]^{2}}}\label{eq:snrn02}
\end{equation}
with this notation.

\begin{figure}[H]
\centering
\includegraphics[width=84mm]{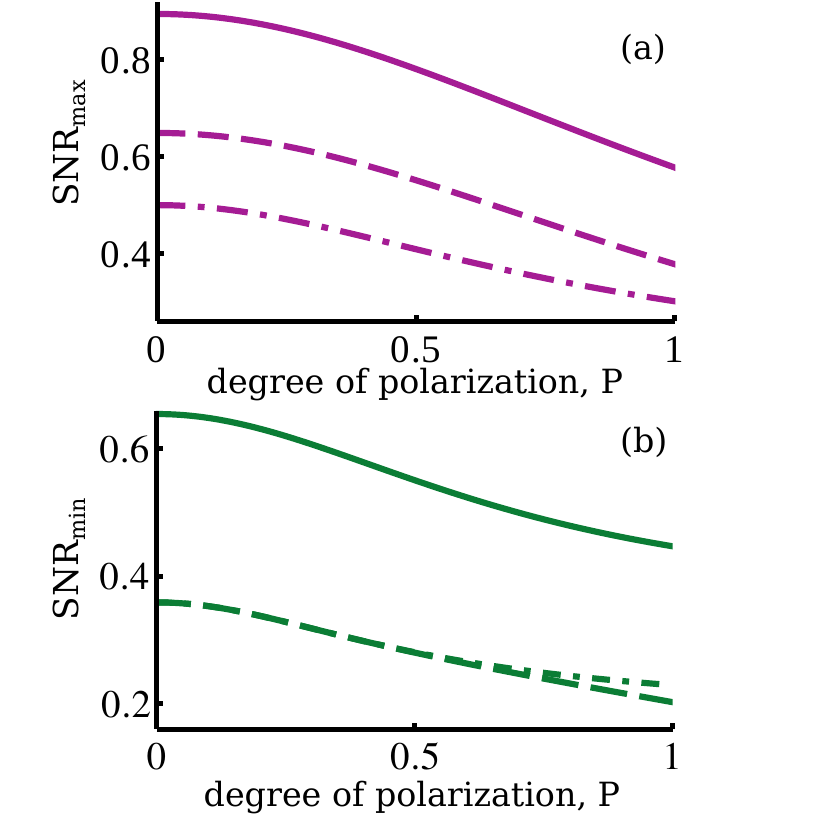}
\caption{\small Dependence of the double- and triple-intensity SNRs on the degree of polarization.
The solid lines correspond to the extremal SNRs of the second-order ghost-imaging setups shown in Fig.~\ref{fig:double}.
The dashed lines depict the maximum and minimum SNRs of the general third-order correlation imaging
arrangement [Fig.~\ref{fig:triple}(a)], while the dash-dotted lines correspond to the extremal SNR values
in the specific triple-intensity ghost-imaging setup [Fig.~\ref{fig:triple}(b)]. The maximum SNRs are
plotted in purple and the minimum SNRs in green. \label{fig:snr}}
\end{figure}

\subsection{SNR in double-intensity imaging}
\vspace{-1ex}

In Sec.~\ref{sec:Foundations-and-geometries} we calculated the second-
and third-order ICFs. To find the SNR in the double-intensity
case we also need the fourth-order ICF $\left\langle I_{1}^{2}I_{2}^{2}\right\rangle $
or its normalized version $\tilde{g}^{(4)}$. Since it is straightforward but tedious, the
computation of the fourth- and higher-order intensity correlations is presented
in Appendix~\ref{sec:Fourth--and-higher-order}.
For a uniformly polarized source and polarization-independent
arms, an exact expression for the SNR as a function of the normalized correlation
$|\hat{\gamma}^{12}|$ and the degree of polarization $P$ is obtained
by inserting Eqs.~\eqref{eq:2icf2e} and \eqref{eq:2x2icf3e-1} into
the definition given in Eq.~\eqref{eq:snrn02}. From this expression we get
the SNR in the dark ($|\hat{\gamma}^{12}|=0$) and light ($|\hat{\gamma}^{12}|=1$) areas of the image.
The double-intensity SNR decreases monotonically as a function of $|\hat{\gamma}^{12}|$,
so it stays within the range
\begin{equation}
\frac{P^{2}+3}{\sqrt{5P^{4}+54P^{2}+21}}\leq\mathrm{SNR}^{(2)}\leq\frac{2}{\sqrt{P^{4}+6P^{2}+5}},\label{eq:snr2max03e2}
\end{equation}
where the upper limit corresponds to the dark areas and the lower
limit to bright areas. Thus, for double-intensity ghost
imaging we have $\mathrm{SNR}_{\mathrm{max}}^{(2)}=\mathrm{SNR}[g_{\mathrm{min}}^{(2)}]$
and $\mathrm{SNR}_{\mathrm{min}}^{(2)}=\mathrm{SNR}[g_{\mathrm{max}}^{(2)}]$,
which are shown with the solid lines in Figs.~\ref{fig:snr}(a)
and \ref{fig:snr}(b), respectively. The SNR is restricted in between these limits
in the absence of other sources of noise. As opposed to the visibility,
the SNR decreases as a function of the DoP. The mathematical reason for the SNR
reduction  with both the correlation parameter $|\hat{\gamma}^{12}|$
and the degree of polarization $P$ is that, although the signal increases
with both $|\hat{\gamma}^{12}|$ and $P$, the level of noise grows proportionally
more due to the stronger dependence of the higher-order intensity correlations
on $|\hat{\gamma}^{12}|$ and $P$ {[}cf.~Eqs.~\eqref{eq:2icf2e}
and \eqref{eq:2x2icf3e-1}{]}. Physically this is explained as follows: when $P$ increases, the source
noise grows as discussed above, and when $|\hat{\gamma}^{12}|$ increases, the random output signals
become more correlated and hence the fluctuations of their product grow.

\subsection{SNR in triple-intensity imaging}
\vspace{-1ex}

Referring to the general triple-intensity ghost-imaging setup of Fig.~\ref{fig:triple}(a),
we can obtain the associated SNR
\begin{equation}
\mathrm{SNR}^{(3)}=\frac{g^{(3)}}{\sqrt{\tilde{g}^{(6)}-\left[g^{(3)}\right]^{2}}}\label{eq:snr301}
\end{equation}
as a function of the magnitudes
of the correlations between any two arms $|\hat{\gamma}^{\alpha\beta}|$
($\alpha,\beta\in\{1,2,3\}$, $\alpha\neq\beta$), the phase difference
between the arms $\Delta\phi_{123}$, and the degree of polarization
$P$, using Eqs.~\eqref{eq:3icf2e1} and \eqref{eq:3x2icf3e}.

First, we find the maximum and minimum signal-to-noise ratios $\mathrm{SNR}_\mathrm{max}^{(3)}$ and
$\mathrm{SNR}_\mathrm{min}^{(3)}$ of the ghost image for each degree of polarization $P$. This is done computationally
by maximizing and minimizing $\mathrm{SNR}^{(3)}$ from Eq.~\eqref{eq:snr301} at each $P$ within the parameter ranges $0\leq|\hat{\gamma}^{\alpha\beta}|\leq1$ and $-1\leq\cos(\Delta\phi_{123})\leq1$.
Next the SNRs in the dark and bright areas of the image, $\mathrm{SNR}[g_\mathrm{max}^{(3)}]$ and
$\mathrm{SNR}[g_\mathrm{min}^{(3)}]$, are obtained with the assumptions
$|\hat{\gamma}^{\alpha\beta}|_{\mathrm{min}}=0$ and $|\hat{\gamma}^{\alpha\beta}|_{\mathrm{max}}=1$ and $\cos(\Delta\phi_{123})=1$, respectively. As in double-intensity correlation imaging, the dark
regions of the image are found to have the maximum possible SNR, i.e., $\mathrm{SNR}[g_\mathrm{min}^{(3)}] = \mathrm{SNR}_\mathrm{max}^{(3)}$, 
shown by the dashed line in Fig.~\ref{fig:snr}(a). However, in the general case the SNR related to the light regions
($\mathrm{SNR}[g_\mathrm{max}^{(3)}]$) is slightly higher than $\mathrm{SNR}_\mathrm{min}^{(3)}$, 
illustrated by the dashed line in Fig.~5(b). In fact, $\mathrm{SNR}[g_\mathrm{max}^{(3)}]$ of the general case 
equals $\mathrm{SNR}_\mathrm{min}^{(3)}$ of the specific case (discussed below), 
shown with the dash-dotted line in the same figure.

The decrease of $\mathrm{SNR}^{(3)}$ when the correlation parameters and the DoP
become larger is due to increased noise, as discussed in connection with the double-intensity case. Physically, $\mathrm{SNR}^{(3)}$
is smaller than $\mathrm{SNR}^{(2)}$ since the product of three partially correlated signals is noisier
than the product of two such signals.

\paragraph{Triple-intensity imaging with one object arm}

The SNR for the specific ghost-imaging setup of Fig.~\ref{fig:triple}(b) is obtained
by inserting Eqs.~\eqref{eq:3icf2e2} and \eqref{eq:3x2icf4e} into Eq.~\eqref{eq:snr301}.
Similarly to the second-order case, the SNR decreases monotonically as a function of $|\hat{\gamma}^{13}|$,
and thus its maximum and minimum values correspond to the SNR in the dark ($|\hat{\gamma}^{13}|=0$) and light ($|\hat{\gamma}^{13}|=1$) regions of the image, respectively.
The SNR extrema are depicted in Fig.~\ref{fig:snr} by dash-dotted lines.
We also found that the SNR in the bright areas of the specific ghost-imaging setup is the same
as that of the general triple-intensity correlation imaging case, while the SNR in the dark areas
of the specific setup is lower than that of the general case.

\begin{figure}[H]
\centering
\includegraphics[width=84mm]{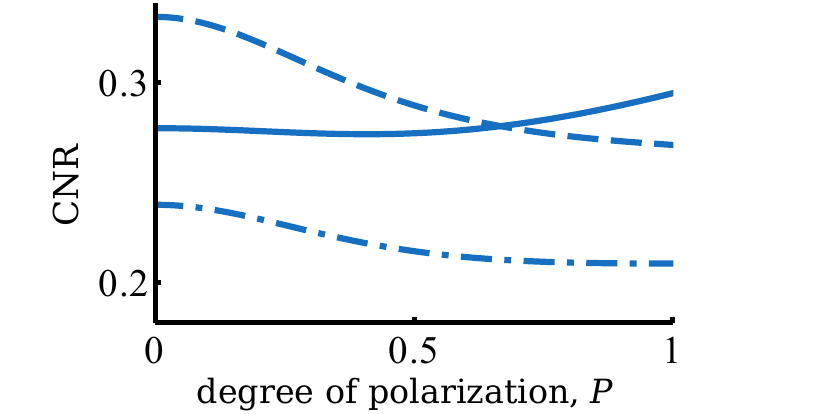}
\caption{\small Dependence of the CNR on the degree of polarization. The solid
line is for the double-intensity case [Eq.~\eqref{eq:2cnr02}],
while the dashed and dash-dotted lines correspond to the CNRs of the
triple-intensity cases related to Fig.~\ref{fig:triple}(a)
[Eq.~\eqref{eq:3cnr02}] and Fig.~\ref{fig:triple}(b)
[Eq.~\eqref{eq:3cnr03}],
respectively. \label{fig:cnr}}
\end{figure}

\section{Contrast-to-noise ratio \label{sec:CNR}}
\vspace{-2ex}

The visibility measures the image contrast
and the SNR assesses the noise in various parts of the
ghost image. To study the contrast compared to the overall noise in the
image we consider a third quality parameter, namely the contrast-to-noise
ratio (CNR). It is defined as \cite{Chan10,Brida11}
\begin{equation}
\mathrm{CNR}^{(N)}\equiv\frac{\left\langle I_{1}\cdots I_{N}\right\rangle _{\mathrm{max}}-\left\langle I_{1}\cdots I_{N}\right\rangle _{\mathrm{min}}}{\sqrt{\frac{1}{2}\left[\mathrm{noise}_{\mathrm{max}}^2\left(I_{1}\cdots I_{N}\right)+\mathrm{noise}_{\mathrm{min}}^2\left(I_{1}\cdots I_{N}\right)\right]}}.\label{eq:ncnr01}
\end{equation}
Referring to Fig.~\ref{fig:gaussian}, we see that the numerator
in Eq.~\eqref{eq:ncnr01} is related to the difference between the
average signal in the bright and dark areas of the image and the denominator
is the root-mean-squared average of the corresponding noises.

\subsection{CNR in double-intensity imaging}
\vspace{-1ex}

The CNR of double-intensity ghost imaging becomes
\begin{equation}
\mathrm{CNR}^{(2)}=\frac{g_{\mathrm{max}}^{(2)}-g_{\mathrm{min}}^{(2)}}{\sqrt{\frac{1}{2}\left\{ \tilde{g}_{\mathrm{max}}^{(4)}-\left[g_{\mathrm{max}}^{(2)}\right]^{2}+\tilde{g}_{\mathrm{min}}^{(4)}-
\left[g_{\mathrm{min}}^{(2)}\right]^{2}\right\} }},\label{eq:2cnr01}
\end{equation}
where we have first substituted from Eq.~\eqref{eq:noise01} and then normalized all
the intensity correlation functions. Using the limits $|\hat{\gamma}^{12}|{}_{\mathrm{min}}=0$
and $|\hat{\gamma}^{12}|{}_{\mathrm{max}}=1$
for the normalized ICFs given in Eqs.~\eqref{eq:2icf2e} and
\eqref{eq:2x2icf3e-1}, we obtain the expression
\begin{alignat}{1}
\mathrm{CNR}^{(2)} & =\frac{P^{2}+1}{\sqrt{3P^{4}+30P^{2}+13}}.\label{eq:2cnr02}
\end{alignat}
For fully polarized light the CNR achieves its maximum, $\mathrm{CNR}_{\mathrm{max}}^{(2)}=\sqrt{2/23}\approx0.29$.
The CNR as a function of $P$ is represented with the solid line
in Fig.~\ref{fig:cnr}. It is almost constant with a slight rise for higher degrees of polarization.

\subsection{CNR in triple-intensity imaging}
\vspace{-1ex}

The triple intensity CNR can be obtained analogously to the second-order case from Eq.~\eqref{eq:ncnr01} and it is
\begin{equation}
\mathrm{CNR}^{(3)}=\frac{g_{\mathrm{max}}^{(3)}-g_{\mathrm{min}}^{(3)}}{\sqrt{\frac{1}{2}\left\{ \tilde{g}_{\mathrm{max}}^{(6)}-\left[g_{\mathrm{max}}^{(3)}\right]^{2}+\tilde{g}_{\mathrm{min}}^{(6)}-\left[g_{\mathrm{min}}^{(3)}\right]^{2}\right\} }}.\label{eq:3cnr01}
\end{equation}
For the general triple-intensity imaging case {[}Fig.~\ref{fig:triple}(a){]}, the normalized ICFs can be retrieved
from Eqs.~\eqref{eq:3icf2e1} and \eqref{eq:3x2icf3e}. Then, inserting
$\cos(\Delta\phi_{123})=1$ and $|\hat{\gamma}^{\alpha\beta}|{}_{\mathrm{max}}=1$ for $g_{\mathrm{max}}^{(3)}$ and
$\tilde{g}_{\mathrm{max}}^{(6)}$, and $|\hat{\gamma}^{\alpha\beta}|{}_{\mathrm{min}}=0$ for $g_{\mathrm{min}}^{(3)}$ and
$\tilde{g}_{\mathrm{min}}^{(6)}$, with ($\alpha,\beta\in\{1,2,3\}$, $\alpha\neq\beta$), the
triple-intensity CNR with respect to the DoP is
\begin{alignat}{1}
\mathrm{CNR}^{(3)} & =\frac{12P^{2}+8}{\sqrt{91P^{6}+1827P^{4}+3033P^{2}+577}}.\label{eq:3cnr02}
\end{alignat}
 Equation \eqref{eq:3cnr02} is shown with the dashed
line in Fig.~\ref{fig:cnr}. For weakly polarized light the improved visibility in third-order
correlation imaging leads to CNR values that are slightly higher as compared to the second-order case.
For strongly polarized light the triple-intensity CNR falls below $\mathrm{CNR}^{(2)}$ due to the
increased noise. The maximum triple-intensity CNR is obtained with completely unpolarized light and it is $\mathrm{CNR}_{\mathrm{max}}^{(3)}=8/\sqrt{577}\approx0.33$.

\paragraph{Triple-intensity setup with one object arm }

As for the general third-order setup, the CNR of the ghost imaging
apparatus with two reference arms and one arm containing the object,
depicted in Fig.~\ref{fig:triple}(b), is given by Eq.~\eqref{eq:3cnr01}.
Inserting Eqs.~\eqref{eq:3icf2e2} and \eqref{eq:3x2icf4e} related to the present setup, the
CNR becomes
\begin{alignat}{1}
\mathrm{CNR}^{(3)} & =\frac{\sqrt{2}\left(5P^{2}+3\right)}{\sqrt{48P^{6}+947P^{4}+1602P^{2}+315}},\label{eq:3cnr03}
\end{alignat}
where we have used the extrema $|\hat{\gamma}^{13}|_{\mathrm{min}}=0$
and $|\hat{\gamma}^{13}|_{\mathrm{max}}=1$ for the correlation between
the reference arms and the test arm. Equation~\eqref{eq:3cnr03}
is shown in Fig.~\ref{fig:cnr} with the dash-dotted line. It behaves otherwise similarly to
the general triple-intensity case {[}Eq.~\eqref{eq:3cnr02}{]} but
is smaller for all $P$. This is because the contrast is lower (Fig.~\ref{fig:vis})
and the mean noise is slightly higher (Fig.~\ref{fig:snr}) for the specific setup
than for the general case. The
maximum CNR is $\mathrm{CNR}_{\mathrm{max}}^{(3)}=\sqrt{2/35}\approx0.24$,
valid for completely unpolarized light.

\section{Summary and conclusions\label{sec:Summary-and-conclusion}}
\vspace{-2ex}

We have taken into account the electromagnetic nature of light and studied the influence
the degree of polarization (DoP) has on the performance of double- and triple-intensity
correlation-imaging arrangements. The illumination was a partially polarized, spatially
incoherent beam of light obeying Gaussian field statistics. Both in the double- and triple-intensity
setups a general intensity-correlation-imaging system and a specific ghost-imaging arrangement
were considered. In all cases the image was assessed in terms of three different
quality parameters, namely the visibility (contrast) and the signal-to-noise (SNR) and
contrast-to-noise (CNR) ratios.

We found that in all setups the image visibility improves when the degree of polarization
increases and it is higher in third-order imaging than in the second-order case.
In contrast, the SNR was found to decrease with increasing degree of polarization
as the source noise grows with the DoP. In addition, the SNR is lower for third-order
imaging than for second-order imaging since in the triple-intensity case the image
is created by a product of a larger number of noisy output signals. The CNR
stays approximately the same as a function of the DoP, although it does go down slightly for
the third-order imaging setups. 
If the CNR is considered as the most important quality parameter, as many authors do, then our results put into question
the earlier suggestions of improved image quality with increased imaging order and DoP. Our results imply
that increasing the imaging order might even be harmful and, on the other hand, changing the DoP does not
have much of an effect and hence no polarizers are necessarily needed in experiments.

The SNR and CNR in ghost imaging with Gaussian light
generally assume low values due to the noisy source. The 
impediments of the source noise
are further enhanced by the correlations and the number of the output fields. The quality
parameters studied in this work do not depend on the state of polarization of the polarized
part of the illuminations, since the arms in the ghost-imaging setups were taken to be
polarization independent.

\section*{Acknowledgments}
\vspace{-2ex}

This research was supported by the Academy of Finland (projects 128331 and 135030) 
and the Japan Society for the Promotion of Science (KAKENHI 23656054).

\numberwithin{equation}{section}

\appendix

\section{Traces of \texorpdfstring{$\mathbf{J}_{0}$}{J0}\label{sec:Traces-of}}
\vspace{-2ex}

In terms of the eigenvalues $J_{1}$ and $J_{2}$ of $\mathbf{J}_{0}$ we have
\begin{alignat}{2}
\mathrm{tr}\left(\mathbf{J}_{0}^{n}\right) & =J_{1}^{n}+J_{2}^{n} &  & =J_{1}^{n}(1+x^{n}),\label{eq:trnjii}\\
\left(\mathrm{tr}\mathbf{J}_{0}\right)^{n} & =\left(J_{1}+J_{2}\right)^{n} &  & =J_{1}^{n}(1+x)^{n},\label{eq:trjiin}
\end{alignat}
 where $x\equiv J_{2}/J_{1}\leq1$. Using the degree of polarization
{[}Eq.~\eqref{eq:p1}{]} we can write
\begin{equation}
x=\frac{1-P}{1+P}.\label{eq:x1}
\end{equation}
This implies that
\begin{equation}
\frac{\mathrm{tr}\left(\mathbf{J}_{0}^{n}\right)}{\left(\mathrm{tr}\mathbf{J}_{0}\right)^{n}}=\frac{\left(1+P\right)^{n}+\left(1-P\right)^{n}}{2^{n}},\label{eq:trndiv}
\end{equation}
where we used Eqs.~\eqref{eq:trnjii}--\eqref{eq:x1}.

\section{Partially coherent source\label{sec:Partially-coherent-source}}
\vspace{-2ex}

To calculate a value for the correlation coefficient \foreignlanguage{british}{$\hat{\gamma}^{12}$} in Sec.~\ref{sub:Triple-intensity} we express the source coherence function as $\mathbf{\Gamma}_{0}(\mathbf{r}_{1}^{\prime},\mathbf{r}_{2}^{\prime})
=\mathbf{J}_{0}\gamma(\mathbf{r}_{2}^{\prime}-\mathbf{r}_{1}^{\prime})$,
where $\gamma(\mathbf{r}_{2}^{\prime}-\mathbf{r}_{1}^{\prime})$ is a normalized
coherence function with the property {$\gamma(\mathbf{0})=1$}.
With the change of variables $\mathbf{r}=\mathbf{r}_{2}^{\prime}-\mathbf{r}_{1}^{\prime}$
and $\mathbf{R}=\left(\mathbf{r}_{1}^{\prime}+\mathbf{r}_{2}^{\prime}\right)/2$,
Eq.~\eqref{eq:Gab3r} becomes
\begin{equation}
\hat{\Gamma}^{\alpha\beta}=\iint\gamma(\mathbf{r})K_{\alpha}^{*}(\mathbf{r}_{\alpha},\mathbf{R}-\frac{\mathbf{r}}{2})K_{\beta}(\mathbf{r}_{\beta},\mathbf{R}+\frac{\mathbf{r}}{2})\mathrm{d}\mathbf{r}\mathrm{d}\mathbf{R}.\label{eq:Gab3rgamma1}
\end{equation}
Inserting the kernels from Eq.~\eqref{eq:klens1} corresponding to the reference arms, the
integral with respect to $\mathbf{R}$ is proportional to $\delta\left[\left(\mathbf{r}_{\beta}-\mathbf{r}_{\alpha}\right)/\left(1-z_{b}/f\right)-\mathbf{r}\right]$
and we find that
\begin{alignat}{1}
\hat{\Gamma}^{\alpha\beta} & =\frac{1}{\left|1-z_{b}/f\right|^{2}}\gamma\left(\frac{\mathbf{r}_{\beta}-\mathbf{r}_{\alpha}}{1-z_{b}/f}\right)\nonumber \\
 & \times\exp\left\{ \frac{ik}{2\zeta}\left[\left(\mathbf{r}_{\beta}^{2}-\mathbf{r}_{\alpha}^{2}\right)\left(1-\frac{z_{a}}{f}\right)-\frac{\left(\mathbf{r}_{\beta}-\mathbf{r}_{\alpha}\right)^{2}}{1-z_{b}/f}\right]\right\} .
\end{alignat}
When {$\mathbf{r}_{1}=\mathbf{r}_{2}$}
the normalized correlation takes on the value
{$\hat{\gamma}^{12}=\hat{\Gamma}^{12}/\sqrt{\hat{\Gamma}^{11}\hat{\Gamma}^{22}}=1$}.

\section{Fourth- and higher-order intensity correlations\label{sec:Fourth--and-higher-order}}
\vspace{-2ex}

To calculate the noise for the SNR and CNR in the double-intensity
case we need to know the fourth-order ICF $\langle I_1^2I_2^2 \rangle$, and for triple-intensity imaging we need
the sixth-order ICF $\langle I_1^2I_2^2I_3^2 \rangle$. These are obtained
from Eqs.~\eqref{eq:nicf1e} and \eqref{eq:nicf1s} with the help of the matrix property
\begin{equation}
\sum_{n_{1},\ldots,n_{m}\in\{x,y\}}A_{n_{1}n_{2}}^{1}A_{n_{2}n_{3}}^{2}\cdots A_{n_{m}n_{1}}^{m}=\mathrm{tr}(A^{1}\cdots A^{m}),
\end{equation}
where
\begin{equation}
A^{i}=\left[\begin{array}{cc}
A_{xx}^{i} & A_{xy}^{i}\\
A_{yx}^{i} & A_{yy}^{i}
\end{array}\right]
\end{equation}
are arbitrary $2\times 2$ matrices for all $i\in\{1,\ldots,m\}$.
Therefore, the fourth-order ICF takes on the form
\begin{alignat}{1}
\left\langle I_{1}^{2}I_{2}^{2}\right\rangle
 & =\left[\mathrm{tr}\mathbf{\Gamma}^{11}\right]^{2}\left[\mathrm{tr}\mathbf{\Gamma}^{22}\right]^{2}
	+\mathrm{tr}(\mathbf{\Gamma}^{11}\mathbf{\Gamma}^{11})\left[\mathrm{tr}\mathbf{\Gamma}^{22}\right]^{2}\nonumber \\
 & +\left[\mathrm{tr}\mathbf{\Gamma}^{11}\right]^{2}\mathrm{tr}(\mathbf{\Gamma}^{22}\mathbf{\Gamma}^{22})
	+\mathrm{tr}(\mathbf{\Gamma}^{11}\mathbf{\Gamma}^{11})\mathrm{tr}(\mathbf{\Gamma}^{22}\mathbf{\Gamma}^{22})\nonumber \\
 & +2\left[\mathrm{tr}(\mathbf{\Gamma}^{12}\mathbf{\Gamma}^{21})\right]^{2}
	+2\mathrm{tr}(\mathbf{\Gamma}^{12}\mathbf{\Gamma}^{21}\mathbf{\Gamma}^{12}\mathbf{\Gamma}^{21})\nonumber \\
 & +4\mathrm{tr}\mathbf{\Gamma}^{11}\mathrm{tr}\mathbf{\Gamma}^{22}\mathrm{tr}(\mathbf{\Gamma}^{12}\mathbf{\Gamma}^{21})
	+4\mathrm{tr}(\mathbf{\Gamma}^{11}\mathbf{\Gamma}^{12}\mathbf{\Gamma}^{21})\mathrm{tr}\mathbf{\Gamma}^{22}\nonumber \\
 & +4\mathrm{tr}\mathbf{\Gamma}^{11}\mathrm{tr}(\mathbf{\Gamma}^{12}\mathbf{\Gamma}^{22}\mathbf{\Gamma}^{21})
	+4\mathrm{tr}(\mathbf{\Gamma}^{11}\mathbf{\Gamma}^{12}\mathbf{\Gamma}^{22}\mathbf{\Gamma}^{21}), \label{eq:2x2icf1e-1}
\end{alignat}
where $\mathbf{\Gamma}^{\alpha\beta}$, with $\alpha,\beta \in (1,2)$, are given by Eq.~\eqref{eq:mcfm1}.
When the source is uniformly polarized and the arms do not depend on polarization,
the normalized version of Eq.~\eqref{eq:2x2icf1e-1} becomes
\begin{alignat}{1}
\tilde{g}^{(4)} & =\frac{P^{4}+6P^{2}+9}{4}+\frac{P^{4}+22P^{2}+9}{2}\bigl|\hat{\gamma}^{12}\bigr|^{2}\nonumber \\
 & +\frac{3P^{4}+10P^{2}+3}{4}\bigl|\hat{\gamma}^{12}\bigr|^{4},\label{eq:2x2icf3e-1}
\end{alignat}
where Eqs.~\eqref{eq:Gab2r} and \eqref{eq:trndiv} were used. Likewise, on normalization and applying
Eqs.~\eqref{eq:Gab2r} and \eqref{eq:trndiv}, the sixth-order ICF associated with the general case of Fig.~\ref{fig:triple}(a)
assumes the form
\begin{alignat}{1}
 \tilde{g}^{(6)} & =\frac{1}{8}\bigl\{(P^{6}+9P^{4}+27P^{2}+27)\nonumber \\
 & +2(P^{6}+25P^{4}+75P^{2}+27)(\left|\hat{\gamma}_{12}\right|^{2}+\left|\hat{\gamma}_{23}\right|^{2}+\left|\hat{\gamma}_{13}\right|^{2})\nonumber \\
 & +(3P^{6}+19P^{4}+33P^{2}+9)(\left|\hat{\gamma}_{12}\right|^{4}+\left|\hat{\gamma}_{13}\right|^{4}+\left|\hat{\gamma}_{23}\right|^{4})\nonumber \\
 & +2(P^{6}+89P^{4}+139P^{2}+27)\nonumber \\
 & \times(\left|\hat{\gamma}_{12}\right|^{2}\left|\hat{\gamma}_{23}\right|^{2}+\left|\hat{\gamma}_{12}\right|^{2}\left|\hat{\gamma}_{13}\right|^{2}+\left|\hat{\gamma}_{13}\right|^{2}\left|\hat{\gamma}_{23}\right|^{2})\nonumber \\
 & +2(P^{6}+33P^{4}+27P^{2}+3)\left|\hat{\gamma}_{12}\right|^{2}\left|\hat{\gamma}_{23}\right|^{2}\left|\hat{\gamma}_{31}\right|^{2}\cos^{2}(\Delta\phi_{123})\nonumber \\
 & +4(3P^{6}+51P^{4}+65P^{2}+9)\nonumber \\
 & \times(\left|\hat{\gamma}_{12}\right|^{2}+\left|\hat{\gamma}_{23}\right|^{2}+\left|\hat{\gamma}_{13}\right|^{2})\left|\hat{\gamma}_{12}\right|\left|\hat{\gamma}_{23}\right|\left|\hat{\gamma}_{31}\right|\cos(\Delta\phi_{123})\nonumber \\
 & +4(P^{6}+57P^{4}+171P^{2}+27)\left|\hat{\gamma}_{12}\right|\left|\hat{\gamma}_{23}\right|\left|\hat{\gamma}_{31}\right|\cos(\Delta\phi_{123})\nonumber \\
 & +2(13P^{6}+117P^{4}+111P^{2}+15)\left|\hat{\gamma}_{12}\right|^{2}\left|\hat{\gamma}_{23}\right|^{2}\left|\hat{\gamma}_{13}\right|^{2}\bigr\}.\label{eq:3x2icf3e}
\end{alignat}
For the specific setup of Fig.~\ref{fig:triple}(b)
we have $|\hat{\gamma}^{12}|=1$, $|\hat{\gamma}^{13}|=|\hat{\gamma}^{23}|$,
and $\Delta\phi_{123}=0$, in which case Eq.~\eqref{eq:3x2icf3e}
reduces to
\begin{alignat}{1}
\tilde{g}^{(6)} & =\frac{3P^{6}+39P^{4}+105P^{2}+45}{4}\nonumber \\
 & +(3P^{6}+111P^{4}+225P^{2}+45)\left|\hat{\gamma}^{13}\right|^{2}\nonumber \\
 & +\frac{15P^{6}+231P^{4}+285P^{2}+45}{2}\left|\hat{\gamma}^{13}\right|^{4}.\label{eq:3x2icf4e}
\end{alignat}
The generation of the sixth-order ICF and its subsequent simplification were handled with symbolic computational
software.

% \bibliographystyle{osajnl}
% \bibliography{Optics17}

\end{multicols}

\end{document}